\documentclass{elsart3}
\usepackage{graphicx}
\usepackage{amsmath}

\newcommand{\phrb}{Phys. Rev. B }

\newcommand{\jap}{J. Appl. Phys. }

\begin{document}

\begin{frontmatter}
\title{Shape and surface anisotropy effects on the hysteresis of ferrimagnetic nanoparticles}

\author{\`{O}scar Iglesias}
\ead{oscar@ffn.ub.es}
\ead[url]{http://www.ffn.ub.es/oscar}
and 
\author{Am\'{\i}lcar Labarta} 
\address{Departament de F\'{\i}sica Fonamental, Universitat de Barcelona, Diagonal 647, 08028 Barcelona, Spain}

\begin{abstract}
We present the results of Monte Carlo simulations of a model of a single maghemite ferrimagnetic nanoparticle with the aim to clarify the role played by the increased anisotropy at the surface and by the shape (spherical or elliptical) of the particle on the magnetization processes at low temperatures. The formation of hedgehog-like structures for high enough surface anisotropy is responsible for a
change in the reversal mechanism of the particles.
\vspace{1pc}
\end{abstract}

\end{frontmatter}

The progressive size reduction of magnetic particles with 
technological application has brought renewed interest on the 
issue of surface and finite-size effects in their magnetic 
properties. 
In particular, it is an old-known fact \cite{Neeljpr54} 
that, when entering the nanometric range, the reduction of 
lattice symmetry at the surface of the particle results in an 
enhancement of magnetic surface anisotropy with respect to 
bulk values which is not clearly understood at present, although
a number of computer simulations have been recently performed 
in order to clarify this issue \cite{First}. 
Here we present the results of Monte Carlo simulations of a 
model of a ferrimagnetic nanoparticle with the spinel 
lattice structure of $\gamma-$Fe$_2$O$_3$ (maghemite). 
We consider an extension of our previous model in 
\cite{Iglesiasprb01a} to the Heisenberg case, including the term 
${ H_{\rm anis}}/k_{B}= -\sum_{i= 1}^{N}
\left(K_C(S_i^z)^2+K_S(\vec{S}_i \cdot \hat n_i)^2 \right)$ 
to account for finite anisotropy.
Here surface spins are allowed to have radial surface anisotropy $K_S$ 
distinct from the uniaxial anisotropy $K_C$ of those in the core. 

By using the standard Metropolis algorithm for continuous 
spins \cite{Hinzkecpc99}, we have computed the spin 
configurations of ellipsoidal particles attained 
after a cooling from a high temperature phase.
We have found that, for high enough $K_S/K_C$ ratios, these 
configurations change from an AF state in which spins in each sublattice 
are almost aligned along the core easy-axis to a 
hedgehog-like structure with spins pointing along radial direction and 
AF order among sublattices (not shown here due to space limitations \cite{WWW}). 
These structures are not observed when uniaxial surface anisotropy 
is considered and therefore can be attributed to the increased 
surface anisotropy. 

\begin{figure}[tbp] 
\centering 
\includegraphics[width=0.35\textwidth]{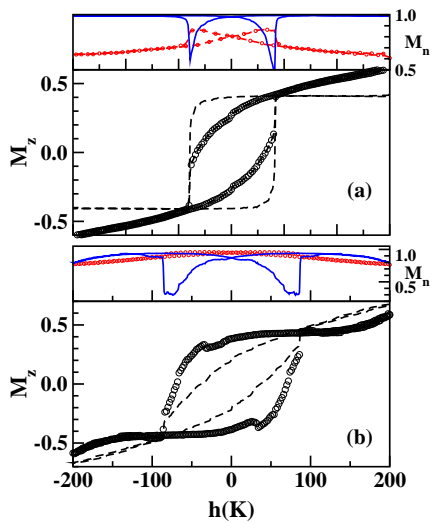}
\caption{Hysteresis loops for a spherical particle with diameter $D=3a$, $K_C=1$, (a) $K_S=10$, (b) $K_S=50$. The main panels show the total magnetization along the z-axis (circles) and the corresponding contribution of the core (dashed lines). The upper subpanels show the surface (circles) and core (continuous lines) contribution to $M_n$ defined in the text. 
}
\label{Fig_1}
\end{figure}

\begin{figure}[tbp] 
\centering 
\includegraphics[width=0.35\textwidth]{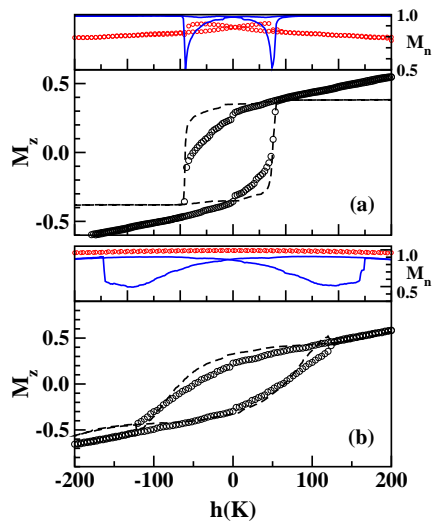}
\caption{Same as in Fig. \ref{Fig_1} but for an elliptical particle with $D=3a$, and long axis $L=8a$.}
\label{Fig_2}
\end{figure}

Zero temperature hysteresis loops for a spherical particle with diameter $D=3a$
and an ellipsoidal particle with the same diameter and major axis $L=8a$
with $K_C=1$ and $K_S= 10, 50$ are compared in Figs. \ref{Fig_1} and \ref{Fig_2}. 
In these figures, we also show the sum of projections surface and core magnetizations into the respective anisotropy directions $M_{n}=\sum_{i=1}^N \left| \vec S_i\cdot \hat n_i \right| $.
First, we notice that as particles become more elongated, although the coercive field and remanence magnetization remain almost constant, the hysteresis loops become more elongated ressembling those observed experimentally. Also the reversal of the core
(dashed line) becomes less uniform, a fact that can be attributed to the increasing proportion of spins with anisotropy direction far from the field direction when increasing $L$.
As $K_S$ increases from $10$ to $50$ the coercive field $h_c$ increases and, more importantly, there is a change in the magnetization reversal mechanism both for spherical and elliptical particles. 
Whereas in the first case the particle core reverses in a quasi-uniform form with the spins pointing mostly along the z-axis ($M_n\approx 1$) during the process except near $h_c$ and with the surface spins following the core reversal (the departure ). 
However, at higher $K_S$ [(b) panels], surface spins remain close to the local radial direction ($M_n\approx 1$) during all the reversal process, driving the core spins out of their local easy axis and making their reversal non-uniform ($0.5<M_n< 1$) due to the appearance of the hedgehog-like structures mentioned above.

%

We acknowledge CESCA and CEPBA under coordination of 
C$^4$ for the computer facilities. This work has been supported by 
SEEUID through project MAT2000-0858 and CIRIT under project 2001SGR00066.

\begin{thebibliography}{10}
\bibitem{WWW}
Snapshots of the configurations will be available at the author's web page.
\end{thebibliography}
%


\end{document}